\documentclass{emulateapj} 

\def\lsim{~\raise0.3ex\hbox{$<$}\kern-0.75em{\lower0.65ex\hbox{$\sim$}}~}
\def\gsim{~\raise0.3ex\hbox{$>$}\kern-0.75em{\lower0.65ex\hbox{$\sim$}}~}

\def\gt{~\hbox{$>$}~}
\def\msun{\rm\,M_\odot}
\def\lbrack2{[\![}
\def\rbrack2{]\!]}

\def\kms{\rm\,km\,s^{-1}}
\def\yr{\rm\,{\rm\,yr}}

\def\myrs{\rm\,{\rm\,Myrs}}
\def\g{\rm\,g}
\def\cm{\rm\,cm}
\def\micron{\rm\,\mu m}

\def\pc{\rm\,pc}
\def\kpc{\rm\,kpc}
\def\mpc{\rm\,Mpc}
\def\mhalo{M_{\rm halo}}

\def\fesc{f_{\rm esc}}
\def\frel{f_{\rm esc}^{\rm rel}}
\def\rvir{r_{\rm vir}}

\def\v90{v_{90}}

\def\vc{v_{\rm c}}
\def\taudust{\tau_{\rm dust}}

\begin{document}

\title{Ionizing radiation from $z=4-10$ galaxies}
\author{Alexei O. Razoumov\altaffilmark{1}}
\author{Jesper Sommer-Larsen\altaffilmark{2,3}}

\shorttitle{Ionizing radiation from $z=4-10$ galaxies} \shortauthors{Razoumov \&
  Sommer-Larsen}

\altaffiltext{1}{Institute for Computational Astrophysics, Department of Astronomy \&
  Physics, Saint Mary's University, Halifax, NS, B3H 3C3, Canada; razoumov@ap.smu.ca}

\altaffiltext{2}{Excellence Cluster Universe, Technische Universit\"at M\"unchen,
  Boltzmannstra\ss e 2, D-85748 Garching, Germany}

\altaffiltext{3}{Dark Cosmology Centre, Niels Bohr Institute, University of
  Copenhagen, Juliane Maries Vej 30, DK-2100 Copenhagen, Denmark;
  jslarsen@astro.ku.dk}

\begin{abstract}
  We compute the escape of ionizing radiation from galaxies in the
  redshift interval $z=4-10$, i.e., during and after the epoch of
  reionization, using a high-resolution set of galaxies, formed in
  fully cosmological simulations. The simulations invoke early,
  energetic feedback, and the galaxies evolve into a realistic
  population at $z=0$. Our galaxies cover nearly four orders of
  magnitude in masses ($10^{7.8}-10^{11.5}\msun$) and more than five
  orders in star formation rates ($10^{-3.5}-10^{1.7}\msun\yr^{-1}$),
  and we include an approximate treatment of dust absorption. We show
  that the source-averaged Lyman-limit escape fraction at $z=10.4$ is
  close to 80\% declining monotonically with time as more massive
  objects build up at lower redshifts. Although the amount of dust
  absorption is uncertain to $1-1.5$ dex, it is tightly correlated
  with metallicity; we find that dust is unlikely to significantly
  impact the observed UV output. These results support reionization by
  stellar radiation from low-luminosity dwarf galaxies and are also
  compatible with Lyman continuum observations and theoretical
  predictions at $z\sim3-4$.
\end{abstract}

\keywords{galaxies: dwarf --- galaxies: formation --- methods: numerical --- stars:
  formation}

\section{Introduction}

The declining number density of quasars at redshift $z\gsim2.5$ \citep{fan-many01}
points to stellar radiation as a likely source of cosmic hydrogen reionization. One
of the key unknown parameters of reionization models is the escape fraction $\fesc$
of ionizing photons from high-redshift galaxies. Observationally, $\fesc$ is much
easier to estimate at lower redshifts. In local starburst galaxies, most studies
arrive at typical values $\fesc\lsim3-10\%$ \citep{leitherer...95, hurwitz..97,
  heckman.....01, bergvall.....06, grimes........07}, corresponding to fairly large
neutral gas covering factors with hydrogen column densities of $\gsim10^{19}\cm^{-2}$
and escape of ionizing radiation through relatively few transparent holes in the
interstellar medium (ISM). Starbursts in the disk of our own Galaxy feature
$\fesc\sim6\%$ normal to the disk \citep{bland-hawthorn.99, bland-hawthorn.01}, a
value which might be a factor of $2-3$ lower when averaged over all directions.

There is strong evidence that the Lyman continuum (LyC) escape fraction evolves with
redshift. A quantity often measured in high-redshift observations is the relative
escape fraction $\frel=\fesc/f_{\rm esc,1500}$ defined as the ratio of the escape
fraction of Lyman-limit photons $\fesc$ to that of $1500{\rm\AA}$ photons. The
conversion between $\fesc$ and the relative escape fractions depends on the amount of
dust attenuation at $1500{\rm\AA}$ within each galaxy and is often uncertain
\citep{inoue....05}. Large observed samples of $z\sim1$ starburst galaxies point to
fairly low $\frel\lsim8-10\%$ \citep{siana-etal07}; however, the ionizing output
changes toward higher redshift. At $z\sim3$, from direct detection of LyC emission
from 2 of the 14 studied star-forming galaxies \citet{shapley....06} put constraints
on the relative escape fraction around $\frel\sim14\%$. More recently,
\citet{iwata.........08} expanded the search to 198 galaxies at $z\approx3.1$,
detecting LyC radiation from seven Lyman break galaxies (LBGs) and 10 Lyman $\alpha$
emitters (LAEs), estimating that prior to accounting for intergalactic medium (IGM)
attenuation the average relative escape fraction is $\frel=61\%$. Taking into account
IGM attenuation with a median LyC opacity $\tau_{\rm IGM}=0.59$ results in
$\frel=110\%$, which in their estimate translates into the average absolute escape
fraction $\fesc=26\%$, and a lower limit of $\fesc=15\%$ prior to accounting for IGM
attenuation. Furthermore, from a comparison of direct observations of the LyC from
galaxies and indirect estimates from intergalactic medium (IGM) ionization,
\citet{inoue..06} found evidence for redshift evolution of the average $\fesc$ from
$1-2\%$ at $z=2$ to $\sim10\%$ at $z\gsim3.6$.

Some of the first theoretical estimates of $\fesc$ were derived from studies of gas
distribution in the Milky Way and from analytical models of galaxy
formation. \citet{dove.94} found $\fesc\sim7\%$ for OB associations in the Galactic
disk. Additional absorption comes from the shells of expanding superbubbles which
further become Rayleigh-Taylor unstable, making the problem of calculating $\fesc$
dependent on time and on details of star formation \citep{dove..00}. Analytical
estimates of $\fesc$ in the cosmological context allowed us to study its dependence
on multiple parameters, such as redshift, dark matter halo mass, baryonic fraction,
star formation efficiency, and stellar ionizing output \citep{ricotti.00, wood.00}.

More accurate theoretical constraints on $\fesc$ can be derived from cosmological
hydrodynamical models incorporating star formation and transfer of ionizing radiation
in the surrounding gas. At very high redshifts ($z\sim20$) estimates show that the
escape fractions reflect the large masses of Population III stars and can easily
approach unity \citep{whalen..04,alvarez..06}. In the observable, post-reionization
universe, there have been several attempts to compute $\fesc$ from cosmological
galaxy formation models. In our earlier simulations \citep{razoumov.06, razoumov.07}
we found that the source-averaged $\fesc$ monotonically declines with redshift, from
$\sim6-10\%$ at $z=3.6$ to $\sim1-2\%$ at $z=2.39$, in line with the observational
findings of \citet{inoue..06}.

\citet{gnedin..08} used a high-resolution simulation of a Lagrangian
region of a $6h^{-1}\mpc$ cosmological model. The selected region
corresponds to five virial radii of a Milky Way-sized galaxy and
includes the Milky Way progenitor and several dozen smaller galaxies
spanning two orders of magnitude in masses and star formation (SF)
rates. Analyzing these galaxies at intermediate ($z=3-5$) redshifts,
\citet{gnedin..08} found that the average escape fractions do not
depend on redshift and tend to correlate positively with the SF rate
dropping below $10^{-3}$ for galaxies with SF rates less than
$\sim1\msun\yr^{-1}$ and dark matter masses below
$\sim10^{10.5}\msun$. Such low $\fesc$ are explained by the fact that
these galaxies have a relatively small gravitational pull and achieve
pressure and density necessary for SF only near the midplane of a
thick neutral disk. The resulting SF regions therefore tend to be
embedded deep inside the opaque disk. Consequently, \citet{gnedin..08}
point out that high-redshift galaxies might be quite inefficient in
emitting ionizing radiation into the IGM.

It is widely agreed that in the absence of energetic supernovae-driven
outflows the increased disk density at higher redshifts $z\sim10$ will
result in very low $\fesc\lsim1\%$ for smooth disks
\citep{wood.00}. On the other hand, simulated galactic disks in cold
dark matter (CDM) cosmologies suffer the excessive loss of angular
momentum resulting in galaxies that are too small compared to
observations \citep{fall.80}. A plausible solution to this angular
momentum problem is heating via feedback from stars and supernovae
that would impede rapid gas cooling and collapse of small
protogalactic gas clouds at high redshifts, allowing gas to preserve a
larger fraction of its angular momentum as it settles into the
disk. \citet{sommer-larsen..03} showed that a population of realistic
disk, lenticular, and elliptical galaxies with almost the correct
amount of angular momentum can be obtained self-consistently in
cosmological simulations by invoking energetic stellar feedback at
early times, either through a much higher SF efficiency than in
present-day galaxies, a more top-heavy initial mass function (IMF), or
both. The affected dwarf galaxies would not have time to settle into a
disk in which feedback would be rather inefficient
\citep{maclow.99}. Instead, they maintain a more spherical or
irregular shape resulting in a much stronger impact from stellar
feedback.

Recently, using very high resolution ($0.1\pc$) coupled radiation hydrodynamics
simulations of dwarf galaxies in the mass range $10^{6.5}-10^{9.5}\msun$ at $z=8$,
\citet{wise.08} showed that the time-averaged $\fesc$ can reach up to 80\% in halos
above $10^8\msun$ with a top-heavy IMF, i.e., in active star-forming galaxies
dominated by the atomic line cooling. It is not clear to what extent $\fesc$ depends
on the IMF as \citet{wise.08} kept the SN feedback strength equal in their normal and
top-heavy IMF calculations varying only the ionizing photon production rate. On the
other hand, lower mass ($<10^{7.5}\msun$), molecular cooling galaxies in their
simulations have much lower SF efficiency and do not contribute a significant flux to
reionize the universe. Irrespective of the mass of the galaxy, \citet{wise.08} found
that $\fesc$ can vary by up to an order of magnitude on the timescale of a few Myrs
which is the dynamical time of a star-forming molecular cloud.

\citet{yajima...09} also found very large escape fractions of order
$20-60\%$ at $z=3.7-7$, analyzing a very high resolution ($1024^3$)
hydrodynamical model of a supernova-dominated primordial galaxy
\citep{mori.06}. However, their model is a single isolated galaxy
taken out of the cosmological context.

In this paper, we take an approach complimentary to that of \citet{wise.08}. Even
though we do not have their ultra-high resolution and we solve the radiative transfer
and ionization equations separately from the equations of hydrodynamics, we examine
the escape of ionizing photons from a set of high-resolution reionization-epoch
galaxies that evolve into a realistic population of galaxies at $z=0$. Specifically,
a two-mode SF prescription is adopted, with both the SF efficiency of the ``early''
mode and the feedback strength calibrated to (1) overcome the angular momentum
problem \citep{sommer-larsen..03}, and to (2) produce a reasonable cosmic enrichment
history that fits observations \citep{sommer-larsen.08}. This efficient high-z
feedback drastically changes the appearance of star-forming galaxies during the
reionization era leading to much higher $\fesc$ than, e.g., in $z=3-5$ dwarf galaxies
\citep{gnedin..08}.

This paper is organized as follows. First, we describe our galaxy
formation models and the technique to compute $\fesc$ in
\textsection\ref{models}. We present the results of our study
including the effect of dust absorption in \textsection\ref{results},
and summarize our findings in \textsection\ref{discussion}.

\section{Model}\label{models}

\subsection{Simulated galaxies}

\begin{figure*}
  \epsscale{1.15}
  \plotone{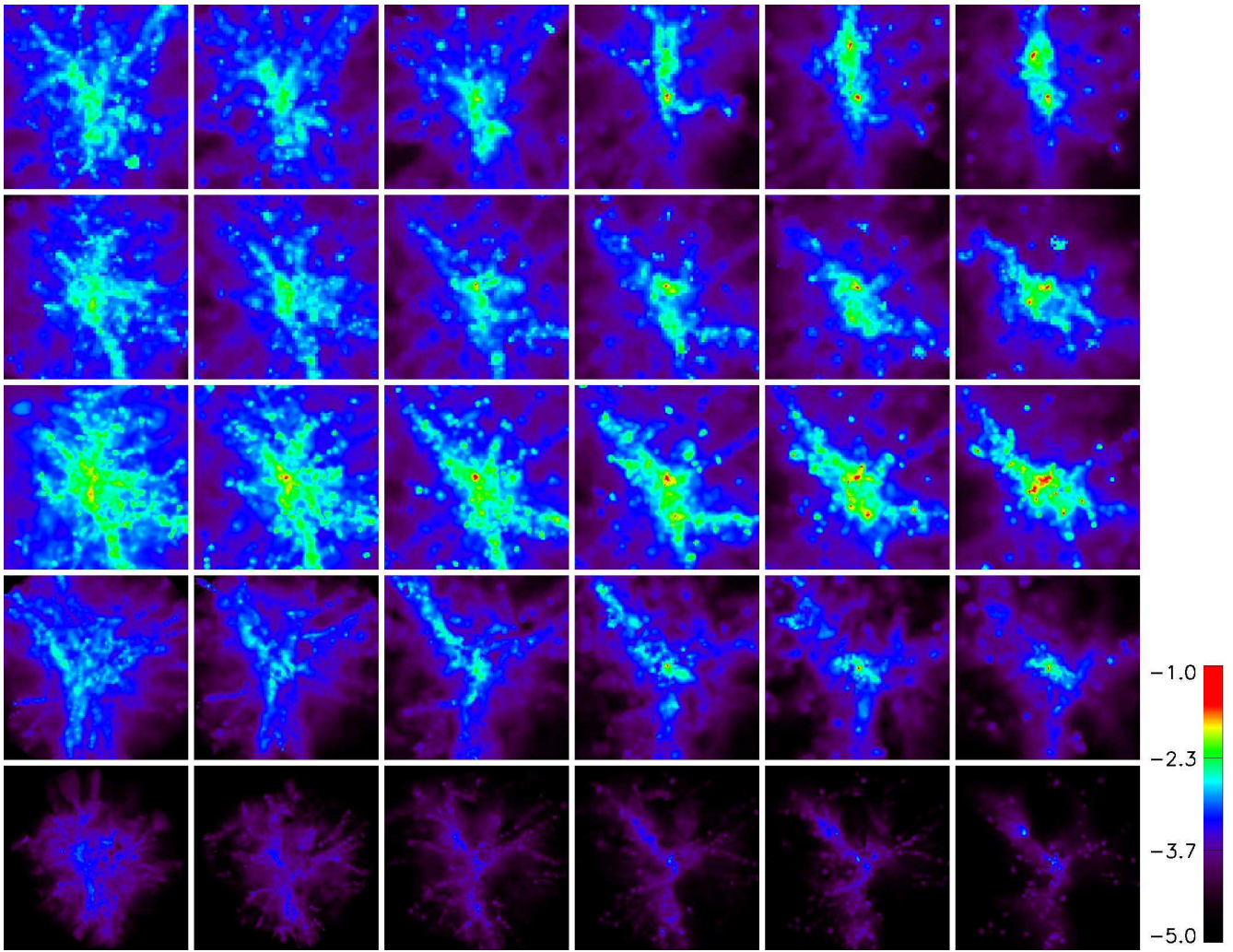}
  \caption{Projected gas density in galaxies S29, S33, S33sc, S108, S108sc (top to
    bottom), at $z=10.4$, 8.2, 6.7, 5.7, 5.0, and 4.4 (left to right). The side of
    the image in each row is 125, 125, 250, 80, 80 physical $\kpc$ (top to
    bottom). The color scale is logarithmic from $10^{-5}\g\cm^2$ to $10^{-1}\g\cm^2$
    in all images.}
  \label{projectedGalaxies}
\end{figure*}

\begin{figure*}
  \epsscale{1.1}
  \plotone{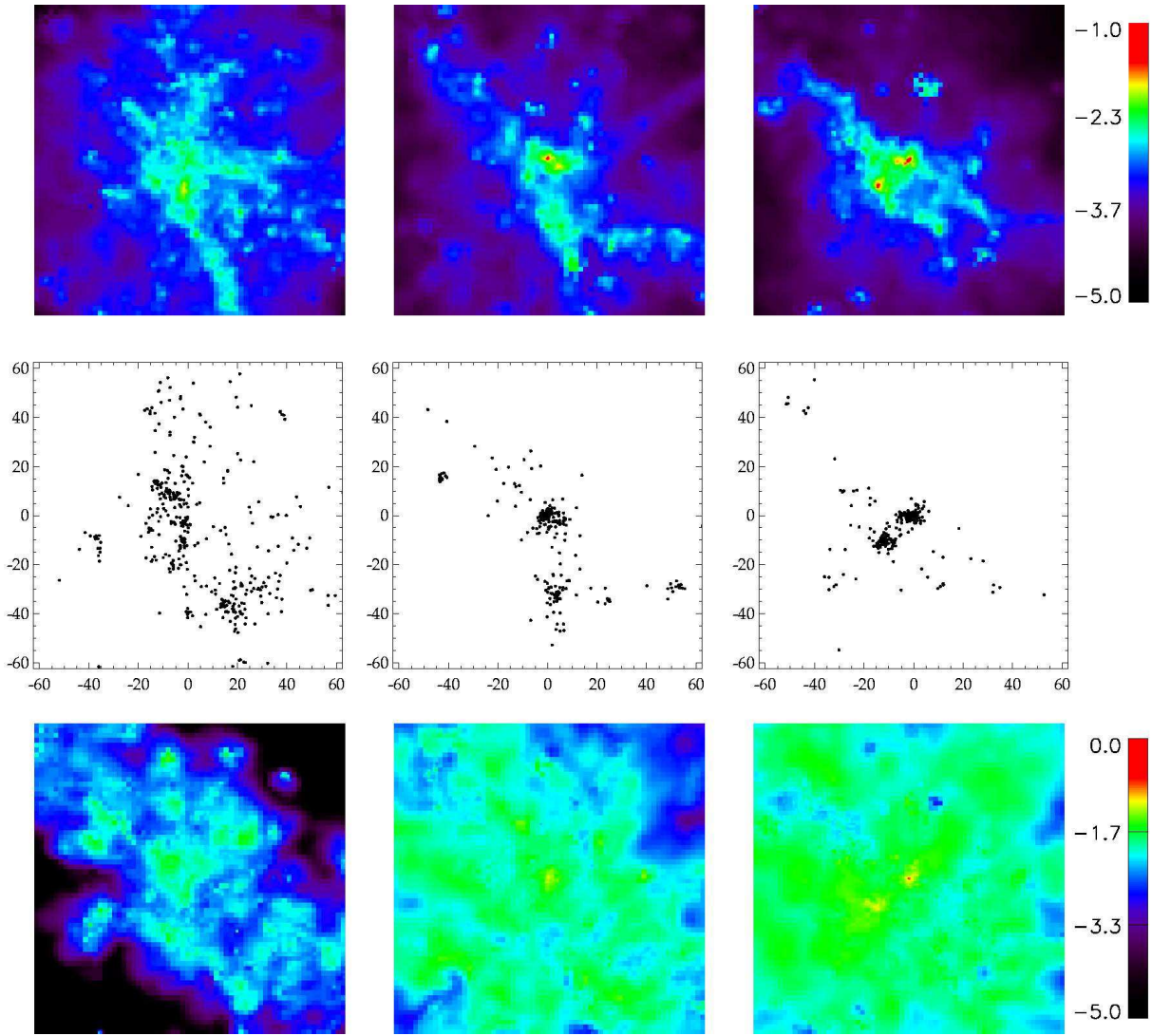}
  \caption{Projected gas density (top panel), stellar particle distribution (center
    panel), and projected mass-weighted oxygen abundance [O/H] (solar units) in
    galaxy S33 at $z=10.4$, 5.7, and 4.4 (left to right). Each image is 125 physical
    $\kpc$ on a side. Both color scales are to the base-10 logarithm, and column
    density in the top panel is measured in $\g\cm^{-2}$.}
  \label{threeDistributions}
\end{figure*}

We use an improved version of the TreeSPH code \citep{sommer-larsen..03} to rerun
nine galaxies selected from a larger cosmological simulation. Table~\ref{galaxies}
lists the following model parameters for each galaxy: physical box size $L$, circular
velocity $\vc$ at $z=0$, mass of an SPH and a dark matter particle, and gravitational
smoothing lengths for SPH and dark matter particles. The code employs the
``conservative entropy'' formulation \citep{springel.02}, includes star formation by
conversion of SPH particles into star particles, and accounts for non-instantaneous
recycling of gas and heavy elements via type II and type Ia SNe and via stellar mass
loss. Chemical evolution is traced using 10 separate elements (H, He, C, N, O, Mg,
Si, S, Ca, and Fe). All galaxies in Table~\ref{galaxies} were computed with the
Salpeter IMF which produces a more reasonable cosmic enrichment history than the
lower metal yield Kroupa IMF \citep{sommer-larsen.08}. Each galaxy was modeled with
at least $10^6$ particles. By $z=0$ these galaxies evolve into systems with a wide
range of masses, from a dwarf galaxy to a super Milky Way sized system,
characteristic of the ``field''. In terms of the luminosity function, they span a
range from super-$L^*$ to about $6-7$ magnitudes below $L^*$. For analysis we use
outputs at $z=10.4$, 8.2, 6.7, 5.7, 5.0, and 4.4 which are separated by
$200\myrs$. Fig.~\ref{projectedGalaxies} shows the evolution of density distribution
in five galaxies that cover a range of masses, and Fig.~\ref{threeDistributions}
displays in addition the distributions of stellar particles and heavy elements in
galaxy S33.

\begin{table}
  \begin{center}
    \caption{List of Modeled Galaxies.}\label{galaxies}
    \begin{tabular}{lcccccc}
      \tableline
      Galaxy & Box Size & $\vc(z=0)$ & $m_{\rm sph}$ & $m_{\rm dm}$ & $\epsilon_{\rm sph}$ &
      $\epsilon_{\rm dm}$\\
      & kpc & $\kms$ & $10^5\msun$ & $10^5\msun$ & kpc & kpc\\
      \tableline

      S29 & 250 & 205 & $1.4$ & $8.1$ & 0.29 & 0.52\\

      S33 & 250 & 180 & $1.4$ & $8.1$ & 0.29 & 0.52\\

      S33sc & 250 & 300 & $8.3$ & $47$ & 0.53 & 0.94\\

      S41 & 250 & 150 & $1.4$ & $8.1$ & 0.29 & 0.52\\

      S87 & 100 & 132 & $0.18$ & $1.0$ & 0.15 & 0.26\\ 

      S108 & 100 & 131 & $0.18$ & $1.0$ & 0.15 & 0.26\\

      S108sc & 80 & 35 & $0.022$ & $0.13$ & 0.07 & 0.13\\ 

      S115 & 100 & 125 & $0.18$ & $1.0$ & 0.15 & 0.26\\ 

      S115sc & 80 & 50 & $0.038$ & $0.22$ & 0.09 & 0.16\\ 

    \end{tabular}
  \end{center}
\end{table}

\subsection{Method}

We model SF by creating discrete star ``particles'', each representing a population
of stars born at almost the same time in accordance with the Salpeter IMF. The
stellar UV luminosity is determined using the population synthesis package Starburst
1999 \citep{leitherer........99} with continuous SF distributed among all stars
younger than $34\myrs$. We compute the effect of this radiation on the surrounding
gas with radiative transfer on top of a nested set of grids containing
three-dimensional distributions of physical variables. To create such a set from the
SPH simulation datasets, we projected each galaxy onto a $128^3$ uniform grid,
divided every base grid cell which contains more than $N_{\rm max}=10$ SPH particles,
and continued this process of subdivision recursively so that no cell contains more
than $N_{\rm max}$ gas particles. For the actual transport of ionizing radiation
around star-forming regions, we employ the ray-tracing scheme with adaptively
splitting radial rays around point sources \citep{abel.02,razoumov.06}. At each
source, we start with 12 isotropic rays using HEALPix discretization on the sphere
\citep{gorski...02} which further split into leaf rays as we move away from the
source or enter a spatially refined region. The local ray angular refinement level
$n_{\rm ray}$ is increased in integer increments to satisfy the condition

\begin{equation}
r_{\rm max}(n_{\rm ray})\ge r_{ij}2^{l_j},
\end{equation}

\noindent
where $r_{ij}$ is the distance between the source $i$ and the cell $j$
of the grid refinement level $l_j$ in which the photoreaction number
and energy rates are computed, and the $r_{\rm max}$ is a
monotonically increasing function chosen to ensure that at least three
ray segments connect every cell in the volume to the each source. The
photoreaction rates are then used iteratively to compute ionization
equilibrium.

In this paper we will work only with the absolute escape fractions
defined as a fraction of photons emitted by the source at a given
energy that reaches the virial radius of the host galaxy. At large
redshifts, star formation is not confined to a few massive galaxies;
instead, it takes place in a large number of smaller galaxies
scattered throughout the volume, and in many cases the concept of a
host halo becomes quite ambiguous. To overcome this problem, for each
star particle we calculate its individual radius of a sphere
containing 200 times the average cosmic critical density at that
redshift which is effectively a local virial radius of that
particle. In calculating the ionization balance, we trace all rays to
the edges of the computational volume; however, $\fesc$ for each
source is evaluated at its individual $\rvir$. To isolate the
contribution to the UVB of sources outside largest halos, we also
identify the most massive halo in each volume and calculate
source-averaged $\fesc$ at its $\rvir$ including only star particles
inside this radius.

\section{Results}\label{results}

\begin{figure}
  \epsscale{1.25}
  \plotone{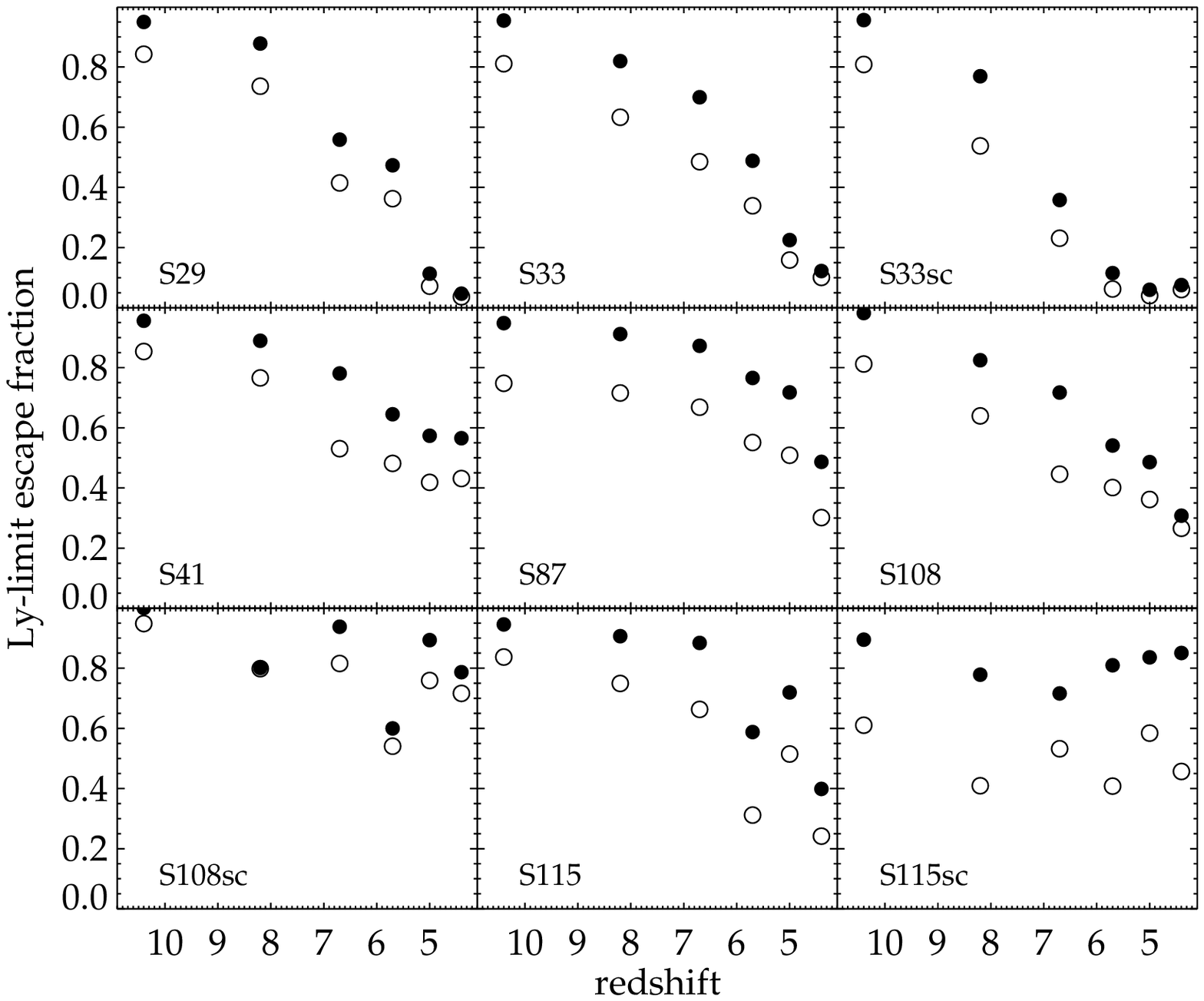}
  \caption{Source-averaged Lyman-limit escape fractions computed at $\rvir$ of the
    most massive halo in each volume (open circles) and at individual $\rvir$ of each
    star particle (filled circles) vs. redshift for the nine galaxies. The effects of
    dust are not included in this plot.}
  \label{redshiftEvolution1}
\end{figure}

\begin{figure}
  \epsscale{1.25}
  \plotone{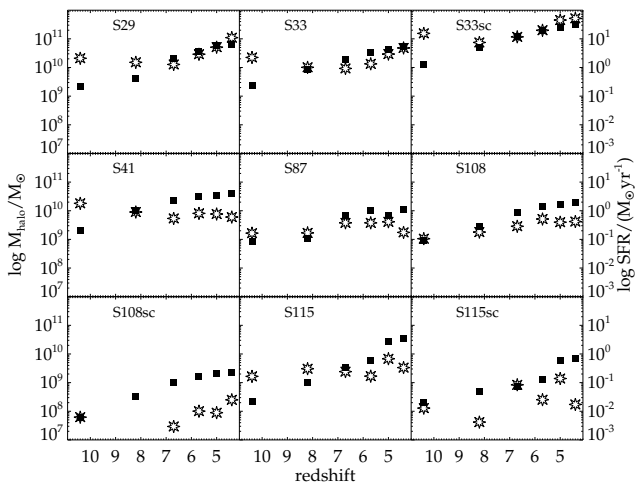}
  \caption{Dark matter halo masses $\mhalo$ (filled squares) and star formation rates
    (star symbols) vs. redshift for the nine galaxies.}
  \label{redshiftEvolution2}
\end{figure}

\begin{figure}
  \epsscale{1.1}
  \plotone{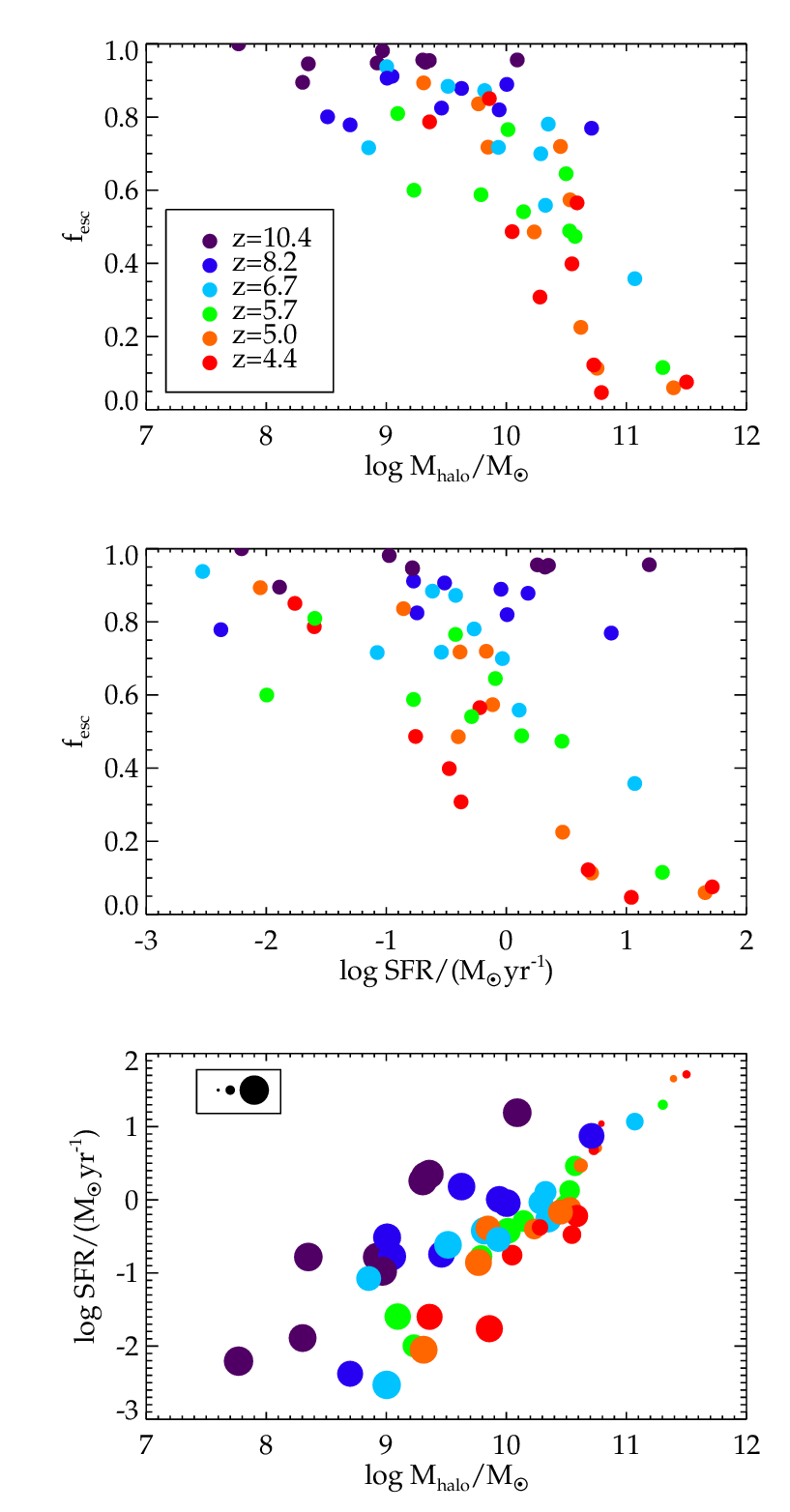}
  \caption{Angular averaged escape fraction as a function of the host
    halo mass (top) and the SF rate (center). Bottom: escape fractions
    in the $\log(\rm SF rate)-\log(M_{\rm halo})$ plane, with the
    radius of each circle proportional to $\fesc^{0.5}$. The insert
    box in the bottom panel shows $\fesc=10^{-2}$, $10^{-1}$, and
    1. Color in all three panels represents redshift with the rainbow
    palette, from $z=10.4$ (violet) to $z=4.4$ (red).}
  \label{mhaloSFRfesc}
\end{figure}

Fig.~\ref{redshiftEvolution1} shows source-averaged Lyman-limit escape fractions of
all galaxies without invoking the effects of dust which will be covered in
\textsection\ref{dustAbsorption}, computed both at the virial radius of the most
massive galaxy and at individual $\rvir$ of each source. There is a clear trend of a
larger $\fesc$ towards higher redshifts. Note that at $z<4.4$ the majority of our
simulated galaxies reproduce the observed gradual decline to $\fesc\sim1-2\%$ at
$z=2.4$ \citep{razoumov.07}. Exceptions are the more massive ($\vc(z=0)>200\kms$)
systems S29, S33 and S33sc which form an early large disk and reach $\fesc<10\%$
before $z=4$. A very interesting result is the large escape fractions of near unity
towards higher redshifts $z\sim8-10$ matching the findings of \citet{wise.08} in
globally turbulent dwarf $10^8-10^{10}\msun$ galaxies at $z=8$. The escape fractions
show an approximate inverse correlation with both the halo mass and the SF rates
(Fig.~\ref{redshiftEvolution2} and Fig.~\ref{mhaloSFRfesc}), coinciding with the
gradual increase in galaxy mass and the decline in SF efficiency with time. Note that
these systems are all actively star-forming galaxies; however, their properties
evolve considerably from $z=10.4$ to 4.4, as they move from the early efficient,
self-propagating SF in turbulent, metal-poor dwarf galaxies to the late inefficient
mode of SF from pre-enriched gas supplied to heavy galactic disks by smooth cooling
flows \citep{sommer-larsen..03}.

Our lower mass ($\sim10^8-10^9\msun$) galaxies are still massive
enough so that even at high SF rates feedback is not efficient at
expelling most of the gas from the galaxy \citep{wise.08} but plays an
important role in shaping small-scale fluctuations in the ISM, leading
to high $\fesc$. This can be seen in all our low-mass, high-SF rate
galaxies all of which have fairly large $\fesc$. On the other hand,
there is a significant spread in properties of individual galaxies as
the escape fraction cannot be derived exclusively from the halo mass
and the SF rate as demonstrated in the lower panel in
Fig.~\ref{mhaloSFRfesc}.

LyC radiation from stars in secondary/satellite galaxies is an important contributor
to the cosmic ionizing background during the early stages of galaxy assembly as
evidenced by the difference between the two types of escape fractions in
Fig.~\ref{redshiftEvolution1} (open and filled circles). On the other hand, in
massive galaxies the overwhelming majority of star-forming regions are found inside
the disks, and the two escape fractions become almost identical.

\begin{figure*}
  \epsscale{1.1}
  \plotone{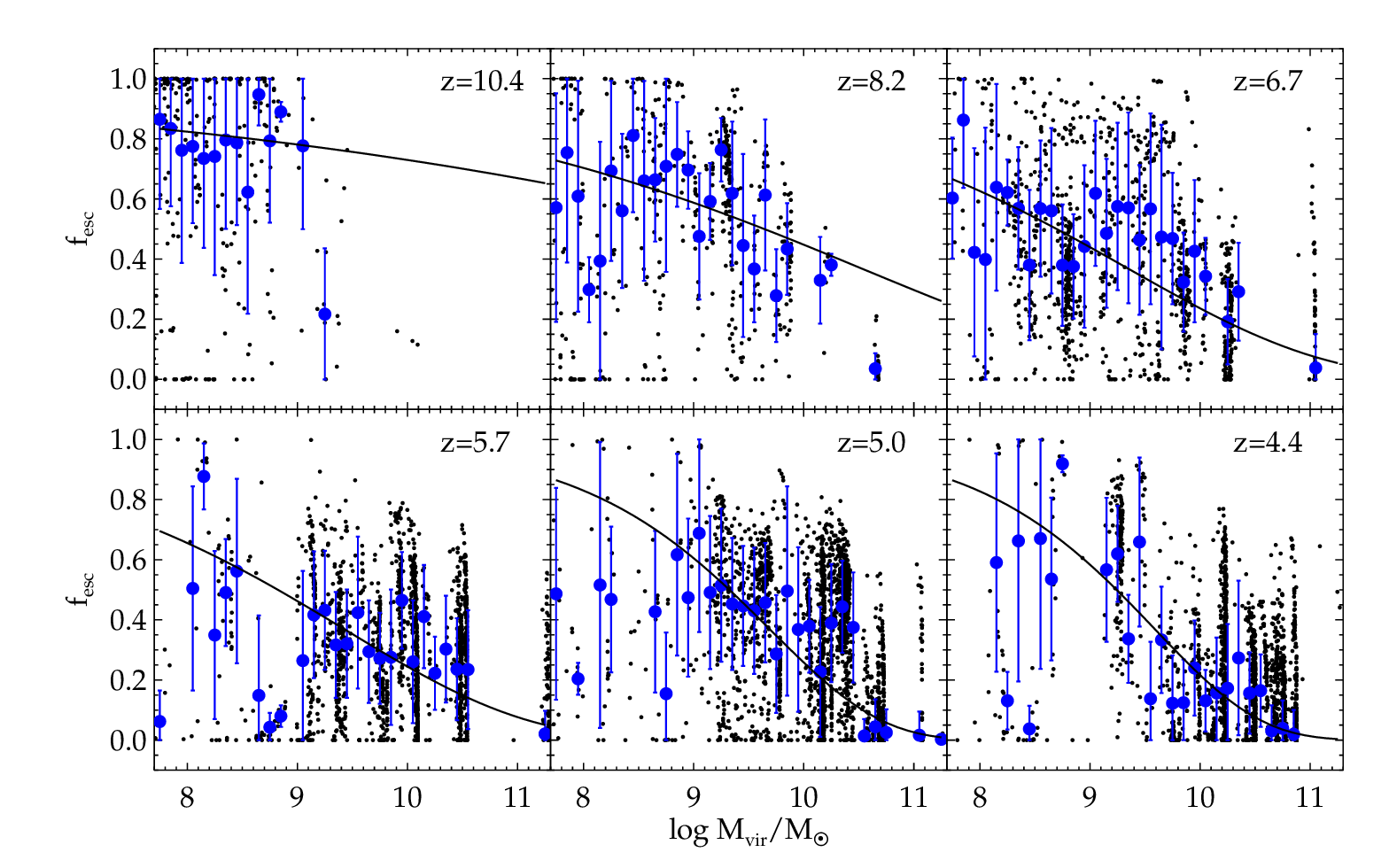}
  \caption{The Lyman-limit escape fractions of all star-forming regions
    (black dots) in all modeled galaxies vs. the host halo mass
    $M(\rvir,z)$ at six redshifts, computed at individual $\rvir$ of
    each star particle. The blue circles and the error bars show
    $\fesc$ averaged inside each mass bin, along with standard
    deviations. The lines show redshift-dependent two-parameter
    exponential fits (see the text for details).}
  \label{individualParticles}
\end{figure*}

In Fig.~\ref{individualParticles}, we use individual star-forming
regions to derive a redshift-dependent fit $\fesc(\mhalo)$. Our
fitting function is based on a simple toy model of the optical depth
$\tau$ to the center of a uniform neutral gas sphere inside the virial
radius $\tau\propto(1+z)^2\mhalo^{1/3}$. We account for deviations
from a uniform sphere -- clumpy gas, non-uniform radial distribution,
partial ionization -- by introducing a second parameter with
$\fesc=\exp(-\alpha\mhalo^\beta)$, where $\alpha$ and $\beta$ are
functions of redshift (Table~\ref{ab}).

\begin{table}
  \begin{center}
    \caption{$\fesc(\mhalo)$ fits.\label{ab}}
    \begin{tabular}{ccc}
      \tableline
      z & $\alpha$ & $\beta$\\
      \tableline
      10.4 & $2.78\times10^{-2}$ & 0.105\\
      8.2  & $1.30\times10^{-2}$ & 0.179\\
      6.7  & $5.18\times10^{-3}$ & 0.244\\
      5.7  & $3.42\times10^{-3}$ & 0.262\\
      5.0  & $6.68\times10^{-5}$ & 0.431\\
      4.4  & $4.44\times10^{-5}$ & 0.454\\
    \end{tabular}
  \end{center}
\end{table}

\subsection{Dust absorption}\label{dustAbsorption}

We follow the prescription of \citet{gnedin..08} to include an approximate treatment
of dust absorption based on the dust extinction curve for the Small Magellanic Cloud
(SMC). We use oxygen abundance as a proxy for metallicity, and assume the SMC
metallicity to be 20\% solar. The dust optical depth can be written as

\begin{equation}
  \taudust(\lambda)=l_{\rm path}n_{\rm H}
  \sigma_{\rm 0, SMC}f(\lambda){Z/Z_\odot\over0.2},
\end{equation}

\noindent
where $l_{\rm path}$ is the physical photon path length, $n_{\rm H}$
is the local neutral or total hydrogen number density, for models with
complete dust sublimation and no sublimation, respectively,
$\sigma_{\rm 0, SMC}=10^{-22}\cm^2$, $Z/Z_\odot$ is oxygen abundance
relative to solar, and $f(\lambda)$ is the parametric
wavelength-dependent fit from \citet{gnedin..08}.

\begin{figure}
  \epsscale{1.1}
  \plotone{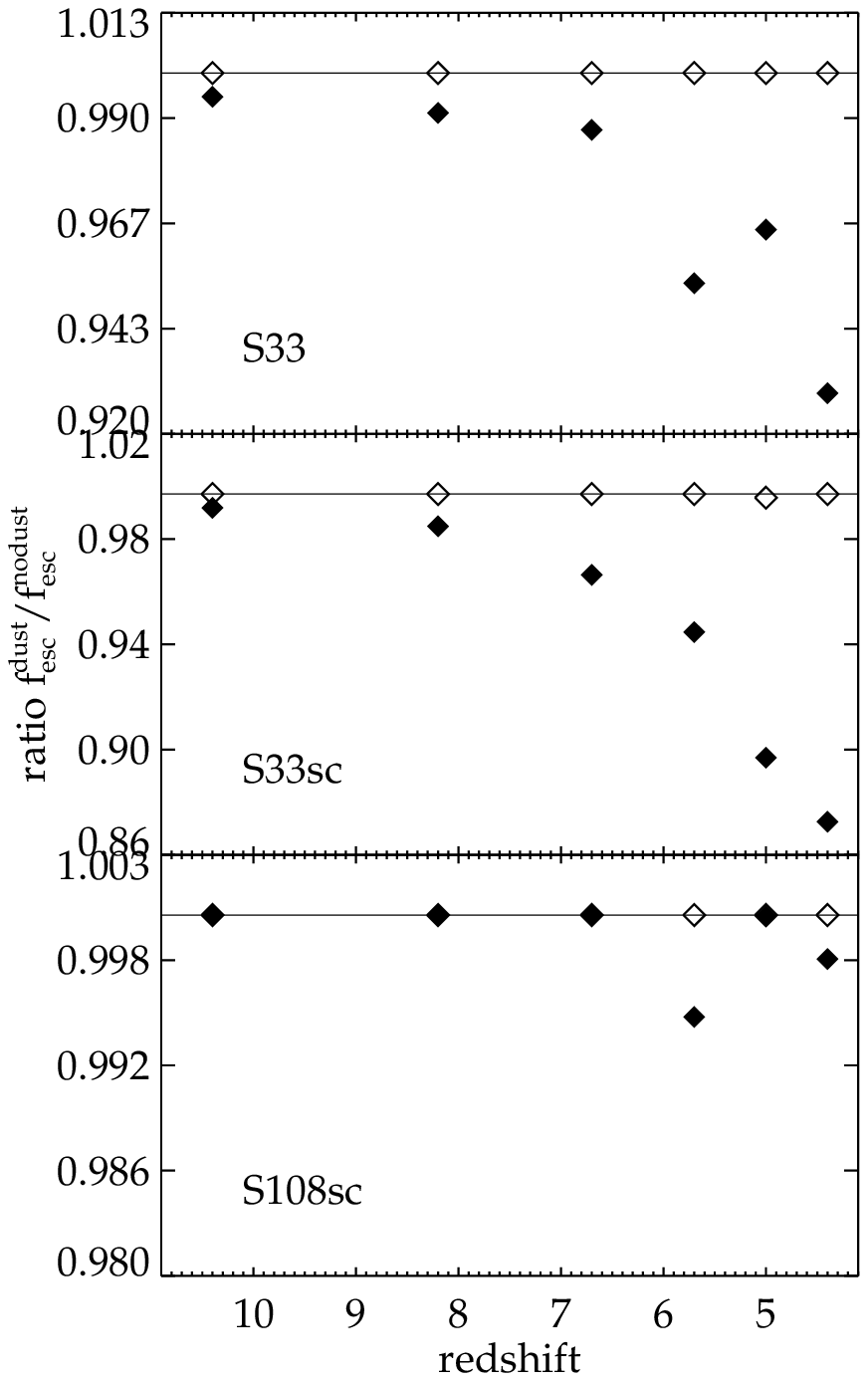}
  \caption{Ratio of Lyman-limit escape fractions with complete dust
    sublimation (open diamonds) and no sublimation (filled diamonds)
    to the escape fractions without dust for two large galaxies (top
    two panels) and a dwarf galaxy (bottom panel), as a function of
    redshift. The solid lines correspond to the ratio of unity.}
  \label{dustEffects}
\end{figure}

Fig.~\ref{dustEffects} shows our results for the two extreme cases, with and without
dust sublimation, for two large and one dwarf galaxies. In the sublimation model
stellar photons destroy dust, and therefore with sufficiently large escape fractions
the effect of dust on ionizing radiation is nil. Only when the escape fraction is
small due to accumulation of neutral hydrogen along the line of sight, we can see
some small effect of dust in the sublimation model, e.g. in the massive galaxy S33sc
at $z=5$, where the relative effect of dust is below 0.2\%. Its impact is more
noticeable in the no-sublimation model. The true escape fractions probably lie in
between these values, as some amount of dust is likely to be destroyed by ionizing
radiation. Similar to \citet{gnedin..08}, we find that the overall effect of dust is
not very large, reducing the escape fractions by few percent at most, explained by
the simple fact that photoionization of neutral hydrogen dominates the optical
depths. Large escape fractions are only observed at low hydrogen column densities,
where dust absorption is usually small. Its effect on the escape of ionizing
radiation is more noticeable in massive galaxies at lower redshifts in which the
buildup of heavy elements leads to relatively high dust concentrations
(Fig.~\ref{dustEffects}).

\begin{figure}
  \epsscale{1.2}
  \plotone{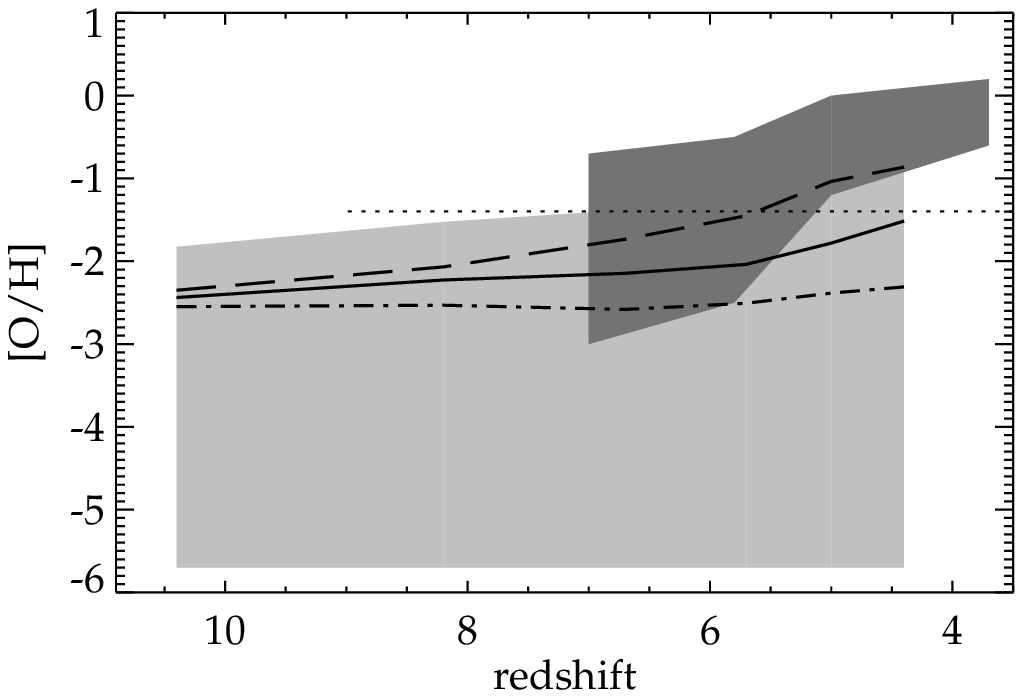}
  \caption{Solid, long-dashed and dash-dotted lines show mass-weighted oxygen
    abundance (relative to solar) in galaxies S33, S33sc, and S108sc as functions of
    redshift. The light-shaded area represents the range of [O/H] containing 90\% of
    the gas mass in S33, i.e., 5\% of the gas has metal enrichment above and below
    this region. The dark-shaded area shows the range of [O/H] in the
    supernova-dominated galaxy from \citet{yajima...09}, and the dotted horizontal
    line corresponds to metallicity adopted in \citet{gnedin..08}.}
  \label{oxygenAbundance}
\end{figure}

It is worth pointing out that the magnitude of dust absorption is highly uncertain
and can range significantly in different studies. Recently, \citet{yajima...09}
suggested another simple model of calculating $\taudust$ from chemical composition
under the following assumptions: (1) absorption is caused by dust grains ranging in
radius from $0.1\micron$ to $10\micron$; (2) the size distribution of dust grains
scales as $r_{\rm d}^{-3.5}$; (3) the density inside dust grains is constant at
$3\g\cm^{-3}$; (4) the diffraction effects are negligible since $r_{\rm
  d}\gt912{\rm\AA}$ ; and (5) the mass density of dust is $0.01Z/Z_\odot\rho_{\rm
  gas}$. The resulting dust absorption is proportional to the total gas density,
similar to the no-sublimation model of \citet{gnedin..08}, but is approximately 25
times weaker than the latter. Despite such low cross-sections, \citet{yajima...09}
find that the escape fractions are significantly regulated by interstellar dust in
their isolated supernova-dominated galaxy. The reason for such a large effect is
their relatively strong metal enrichment producing on average solar metallicity ISM
at $z=3.7$ (Fig.~\ref{oxygenAbundance}) which in addition is very irregular and
moderately optically thin for ionizing photons.

Observations of UV-selected star-forming galaxies at $z\sim2$ point to
metallicities in the range 0.1-0.5 dex below solar
\citep{erb.....06}. If star formation peaks at $z\sim2-3$,
metallicities should be substantially lower in the interval
$z\sim10.4-4.4$ we study here. All models in our set feature a gradual
metal enrichment such that most of the massive galaxies have ${\rm
  [O/H]}\sim0.1-0.5$ at $z=2$ (Fig.~\ref{oxygenAbundance}). Since the
no-sublimation models put an upper limit on dust absorption
(Fig.~\ref{dustEffects}), we conclude that its effect is unlikely to
significantly impact the output of ionizing photons in $z\gsim4.4$
galaxies, except for very rare, unusually massive systems.

\subsection{Resolution effects}

For the numerical convergence study, we use an additional model K33 to
compare the escape fractions at $z=5.85$ obtained at standard (K33-64)
and 8 times the mass resolution (K33-512), while keeping the strength
of stellar feedback and the ionizing luminosity per unit stellar mass
constant. At these two resolutions, this galaxy contains 755 star
particles of mass $1.42\times10^5\msun$, and 8447 star particles of
mass $1.78\times10^4\msun$, respectively.

\begin{figure}
  \epsscale{1.1}
  \plotone{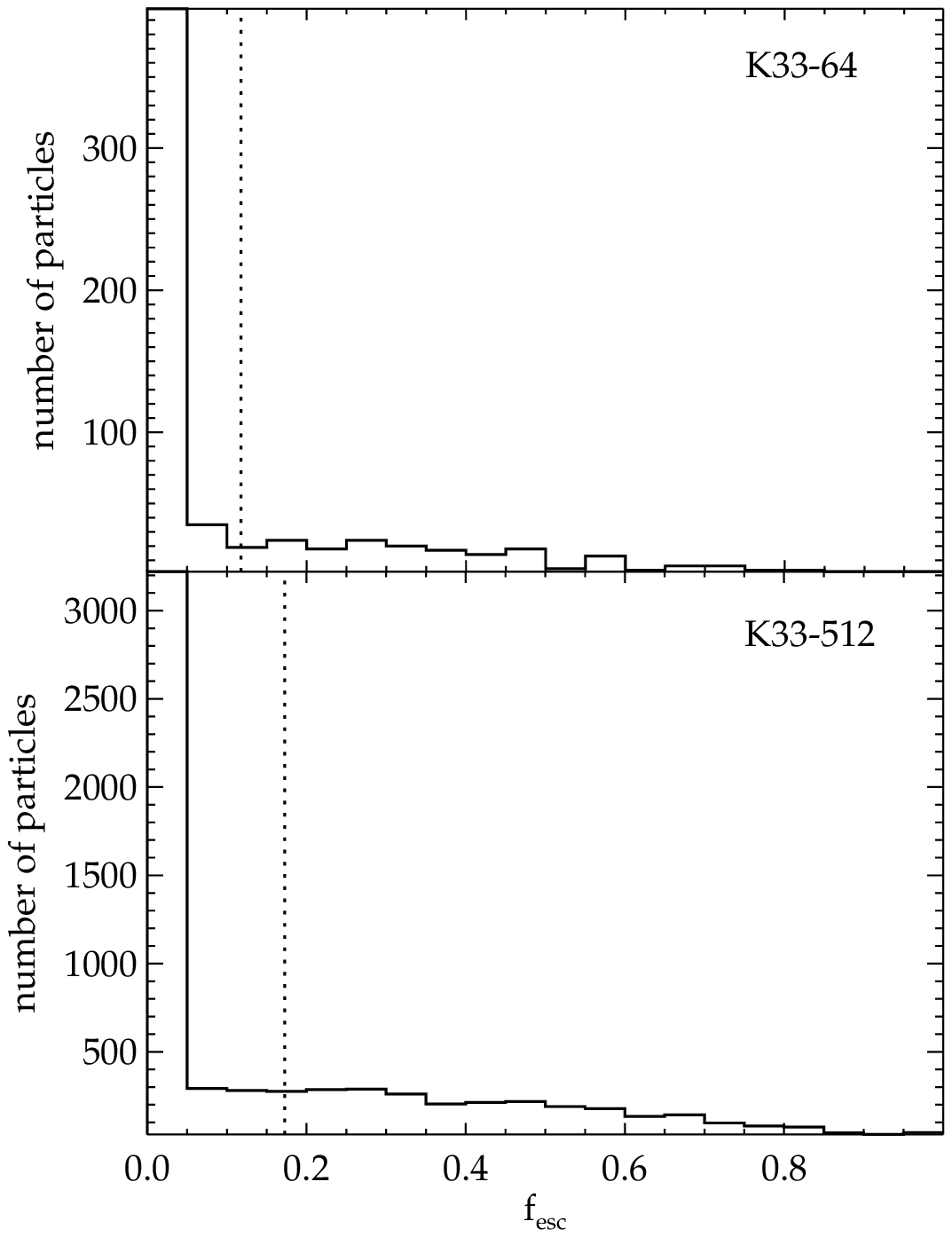}
  \caption{Distribution of star particles by their Lyman-limit $\fesc$ at individual
    particle's $\rvir$ at two different numerical resolutions at $z=5.85$. The dashed
    lines show source-averaged $\fesc$.}
  \label{resolutionStudy}
\end{figure}

We find the Lyman-limit $\fesc=(0.142, 0.118, 0.101)$ for K33-64, and (0.215, 0.173,
0.143) for K33-512, at $r=(0.5,1,2)\times\rvir$, respectively
(Fig.~\ref{resolutionStudy}). The star formation rate in the higher resolution model
is slightly larger, $4.4\msun\yr^{-1}$ versus $3.2\msun\yr^{-1}$, which could explain
the observed difference in $\fesc$, since the two quantities should be coupled
\citep{wise.08}. In addition, we see the increased porosity of the ISM in the higher
resolution model, with more transparent channels between dense clumps through which
radiation can escape. To be conservative, the escape fractions presented in this work
should probably be considered as lower limits.

\section{Discussion and conclusions}\label{discussion}

We have coupled high-resolution galaxy formation models with
point-source radiative transfer to compute the escape fractions of LyC
photons from galaxies at $z=10.4$, 8.2, 6.7, 5.7, 5.0, and 4.4. We
confirm very large escape fractions of near unity at $z\sim8-10$ found
by \citet{wise.08} in dwarf $10^8-10^{10}\msun$ galaxies, conducive to
efficient stellar reionization. The difference between this result and
much lower escape fractions in $\mhalo\gsim8\times10^9\msun$ galaxies
with similar SF rates at $z=3-4$ \citep{gnedin..08} reflects the model
differences between reionization-epoch star-forming galaxies and
lower-redshift dwarf galaxies with a relatively low SF
efficiency. There are several physical factors that could account for
very different SF efficiencies in $z\sim6-10$ galaxies. These systems
are likely to have very low metallicities for which the strength of
stellar winds is greatly reduced \citep{kudritzki02}, producing less
local disruption in star-forming clouds, and leading to higher SF
efficiency. Lower metallicities correspond to less efficient cooling;
however, this effect can be offset by the fact that -- at a fixed halo
mass -- higher redshift galaxies are more compact resulting in more
efficient cooling. In addition, there are theoretical expectations of
a more top-heavy IMF at high redshifts \citep{abel..02, bromm..02,
  padoan.02}. A larger fraction of massive stars is likely to yield
more efficient feedback on galactic scales, producing a larger number
of transparent channels in the ISM through which ionizing radiation
can escape the galaxy.

All our systems are proto-spiral galaxies residing in a ``field'' environment. We did
not consider more crowded proto-elliptical environments in which evolution is
accelerated due to the higher overdensity. In such galaxies the UV escape
fractions could drop quicker with time than what is seen in our current study. In
addition, higher metallicity might amplify the effect of dust although relatively few
UV photons would escape from such environments. We will address more crowded systems
in a future study.

We did not compute the effect of ionizing radiation on the
hydrodynamical flow which is instead shaped by the thermal feedback
energy. Would our results change if we used a coupled
radiation-hydrodynamics approach? Ionizing radiation heats up the gas
which can then expand and leave the star-forming
region. \citet{gnedin..08} found overall that coupling of radiative
transfer and hydrodynamics does not produce a large change in
$\fesc$. For the dwarf galaxies considered by \citet{wise.08}, they
reported fast variations of $\fesc$ by up to an order of magnitude on
the timescale of a few Myrs which is the dynamical timescale of a
star-forming molecular cloud. These changes in $\fesc$ are attributed
to gas expulsion by winds and supernovae creating transparent channels
for ionizing photons. Since feedback by winds and supernovae is
already a key part of our hydrodynamical calculations, we argue that
our postprocessing of fixed time snapshots of galaxies with radiative
transfer with the assumption of ionization equilibrium gives a fairly
accurate estimate of the escape fractions. On the other hand, we
cannot compute $\fesc$ variations on short timescales of few Myrs, and
therefore we cannot tell exactly how representative our computed
$\fesc$ of the time-average escape fractions, something we are hoping
to address in the future.

The other possible source of uncertainty in computing $\fesc$ is gas
clumping on sub-resolution scales. Since for fixed cloud mass and
ionization the optical depth scales as $R^{-2}$ with the cloud radius
$R$, and recombinations are more efficient at higher densities, the
true $\fesc$ could be smaller than our current estimates. On the other
hand, star-forming clouds are shaped by feedback, and the simple
$\tau\propto R^{-2}$ scaling might not be valid. Moreover, higher grid
resolution would yield a more porous ISM, with a strong network of
transparent channels through which radiation can escape.

\section*{Acknowledgments}

We thank Nick Gnedin for providing his convenient parametrization of dust extinction,
and the anonymous referee for many useful comments. Computational facilities for this
work were provided by ACEnet, the regional high performance computing consortium for
universities in Atlantic Canada. We also gratefully acknowledge abundant access to
the computing facilities provided by the Danish Centre for Scientific Computing
(DCSC). ACEnet is funded by the Canada Foundation for Innovation (CFI), the Atlantic
Canada Opportunities Agency (ACOA), and the provinces of Newfoundland \& Labrador,
Nova Scotia, and New Brunswick. AR acknowledges financial support from ACEnet. This
work was supported in part by the DFG Cluster of Excellence "Origin and Structure of
the Universe". The Dark Cosmology Centre is funded by the Danish National Research
Foundation.

\bibliographystyle{apj}


\end{document}